\documentclass{emulateapj}
\slugcomment{Accepted to ApJ: September 8, 2008}
\usepackage{amsmath}

\shorttitle{Discovery of an Unusual Optical Transient}
\shortauthors{Barbary et al.}

\begin{document}

\title{Discovery of an Unusual Optical Transient with the \emph{Hubble Space Telescope}
       \altaffilmark{1}}

\author{
K.~Barbary\altaffilmark{2,3},
K.~S.~Dawson\altaffilmark{3},
K.~Tokita\altaffilmark{4},
G.~Aldering\altaffilmark{3},
R.~Amanullah\altaffilmark{5},
N.~V.~Connolly\altaffilmark{6},
M.~Doi\altaffilmark{4},
L.~Faccioli\altaffilmark{7}, 
V.~Fadeyev\altaffilmark{8},
A.~S.~Fruchter\altaffilmark{9},
G.~Goldhaber\altaffilmark{2,3},
A.~Goobar\altaffilmark{5},
A.~Gude\altaffilmark{2},
X.~Huang\altaffilmark{2},
Y.~Ihara\altaffilmark{4},
K.~Konishi\altaffilmark{10},
M.~Kowalski\altaffilmark{11},
C.~Lidman\altaffilmark{12},
J.~Meyers\altaffilmark{2,3},
T.~Morokuma\altaffilmark{4,13},
P.~Nugent\altaffilmark{3},
S.~Perlmutter\altaffilmark{2,3},
D.~Rubin\altaffilmark{2,3},
D.~Schlegel\altaffilmark{3},
A.~L.~Spadafora\altaffilmark{3},
N.~Suzuki\altaffilmark{3},
H.~K.~Swift\altaffilmark{2,3}, 
N.~Takanashi\altaffilmark{4},
R.~C.~Thomas\altaffilmark{3,14},
N.~Yasuda\altaffilmark{10}
(The Supernova Cosmology Project)
}

\altaffiltext{1}{Based in part on observations made with the NASA/ESA Hubble 
  Space Telescope, obtained from the data archive at the Space Telescope 
  Institute. STScI is operated by the association of Universities for 
  Research in Astronomy, Inc. under the NASA contract NAS 5-26555. 
  The observations are associated with program GO-10496.
  Based in part on observations obtained at the European Southern 
  Observatory under ESO program 077.A-0110.
  Based in part on observations collected at Subaru Telescope, which is
  operated by the National Astronomical Observatory of Japan.
  Some of the data presented herein were obtained at the W.M. Keck Observatory,
  which is operated as a scientific partnership among the California Institute 
  of Technology, the University of California and the National Aeronautics and 
  Space Administration.}

\altaffiltext{2}{Department of Physics, University of California Berkeley, Berkeley, 94720-7300 CA, USA (email: kbarbary@berkeley.edu).}
\altaffiltext{3}{E. O. Lawrence Berkeley National Laboratory, 1 Cyclotron Rd., Berkeley, CA 94720, USA. }
\altaffiltext{4}{Institute of Astronomy, Graduate School of Science, University of Tokyo, 2-21-1 Osawa, Mitaka, Tokyo 181-0015, Japan.}
\altaffiltext{5}{Department of Physics, Stockholm University,  Albanova University Center, S-106 91 Stockholm, Sweden.}
\altaffiltext{6}{Department of Physics, Hamilton College, Clinton, NY 13323, USA.}
\altaffiltext{7}{Space Sciences Laboratory, University of California Berkeley, Berkeley, CA 94720, USA.}
\altaffiltext{8}{Santa Cruz Institute for Particle Physics, University of California, Santa Cruz, CA 95064, USA.}
\altaffiltext{9}{Space Telescope Science Institute, 3700 San Martin Drive, Baltimore, MD 21218, USA.}
\altaffiltext{10}{Institute for Cosmic Ray Research, University of Tokyo,
5-1-5, Kashiwanoha, Kashiwa, Chiba, 277-8582, Japan.}
\altaffiltext{11}{Humboldt Universit\"at Institut f\"ur Physik, Newtonstrasse 15, Berlin 12489, Germany.}
\altaffiltext{12}{European Southern Observatory, Alonso de Cordova 3107, Vitacura, Casilla 19001, Santiago 19, Chile.}
\altaffiltext{13}{National Astronomical Observatory of Japan, 2-21-1 Osawa, Mitaka, Tokyo 181-8588, Japan.}
\altaffiltext{14}{Luis W. Alvarez Fellow.}

\begin{abstract}
We present observations of SCP 06F6, 
an unusual optical transient discovered during 
the \emph{Hubble Space Telescope} Cluster Supernova Survey. 
The transient brightened over a period of $\sim$100 days, 
reached a peak magnitude of $\sim$21.0 in both $i_{775}$ and $z_{850}$, 
and then declined over a similar timescale. 
There is no host galaxy or progenitor star detected at the location of 
the transient to a $3\sigma$ upper limit of $i_{775} \ge 26.4$ and 
$z_{850} \ge 26.1$, giving a corresponding lower limit on the flux increase 
of a factor of $\sim$120. 
Multiple spectra show five broad absorption bands between $4100$~\AA\ and 
$6500$~\AA\ and a mostly featureless continuum longward of $6500$~\AA. 
The shape of the lightcurve is inconsistent with microlensing.
The transient's spectrum, in addition to being inconsistent with all 
known supernova types, is not matched to any spectrum in the 
Sloan Digital Sky Survey (SDSS) database.
We suggest that the transient may be one of a new class.
\end{abstract}


\section{Introduction}

Supernova (SN) surveys are designed to detect the brightening of 
supernovae over timescales of days to weeks. They often cover large areas at 
high sensitivity. As a result, they are able to discover unusual and rare 
transients with similar timescales.
For example, in 2006 the Lick Observatory Supernova Search (LOSS) 
discovered an optical transient in the galaxy M85 
\citep{kulkarni07a,rau07b,ofek08a} with a lightcurve plateau of $\sim$80 days. 
It is suggested that the origin of this transient is a stellar merger and 
that an entire class of similar transients, \emph{luminous red novae}, 
exists.
Other recent discoveries of rare objects include a Type Ia SN with a 
super-Chandrasekhar mass progenitor \citep{howell06a} from the Supernova 
Legacy Survey (SNLS) and SN 2005ap, the most luminous SN ever observed
\citep{quimby07a} from the Texas Supernova Search (TSS).

Here we report the observations of the optical transient SCP 06F6 
discovered 
during the course of the 2005-2006 \emph{Hubble Space Telescope} (\emph{HST}) 
Cluster SN Survey (PI: Perlmutter; Dawson et al. 2008, in preparation). 
The discovery was originally reported in a June 2006 IAU circular 
\citep{dawson06a}.
Its lightcurve rise time of $\sim$100 days is 
inconsistent with all known SN types, and its spectroscopic 
attributes are not readily matched to any known variable. 
We present photometry in \S2 and spectroscopy in \S3. In \S4, 
we discuss constraints and summarize.

\section{Photometry}

The optical transient was discovered on 21 February 2006 (UT dates are used 
throughout this paper) in images taken in the course of the \emph{HST} 
Cluster SN Survey in a field centered on galaxy cluster CL 1432.5+3332.8 
\citep[redshift $z = 1.112$;][]{elston06a}. 
As part of the survey this field was imaged repeatedly over nine epochs 
with the Advanced Camera for Surveys (ACS) Wide Field Camera  
with a cadence of roughly three weeks. Each epoch consisted of four exposures 
in the $F850LP$ filter (hereafter denoted $z_{850}$; similar to SDSS $z'$) 
totaling $\sim$1400~s and one exposure of $\sim$400~s in the $F775W$ filter 
(hereafter denoted $i_{775}$; similar to SDSS $i'$). 
Table~\ref{tb:lightcurve} gives a summary of photometric observations. 
Cosmic ray rejection was performed on the $z_{850}$ images and each 
epoch was searched for SNe using a modified version of the image subtraction 
code developed by the Supernova Cosmology Project \citep{perlmutter99a} 
employing earlier epochs as a reference. 
The transient was discovered in the fourth epoch and is located at 
$\alpha = 14^\mathrm{h} 
32^\mathrm{m} 27^\mathrm{s}.395$, $\delta = +33^\circ 32' 24''.83$ (J2000.0), 
corresponding to galactic coordinates $l = 55.528943^\circ$, 
$b = 67.345346^\circ$ and ecliptic coordinates 
$\lambda = 13^\mathrm{h} 24^\mathrm{m} 9^\mathrm{s}.067$, 
$\beta = 45^\circ 21' 46''.06$.
This position has a statistical uncertainty of $0''.01$ relative to the 
\emph{HST} Guide Star Catalogue 2.3.2, which has an overall systematic 
uncertainty of $0''.3$. 
The angular separation from the cluster center is $35''$, 
corresponding to a projected physical separation at the 
cluster redshift of 290~kpc. 
There is no prior detection of a source at the transient's location in the 
NRAO VLA Sky Survey \citep{condon98a} at 1.4~GHz to the survey 
$5\sigma$ detection limit of $2.5~\mathrm{mJy}~\mathrm{beam}^{-1}$. 
There is no X-ray detection at this location in a 5 ks 
exposure in the Chandra Telescope XBootes survey \citep{kenter05a} to 
the detection limit of $7.8 \times 10^{-15}~\mathrm{erg}~
\mathrm{cm}^{-2}~\mathrm{s}^{-1}$ in the full $0.5$-$7$~keV band. 

\begin{table*}
\begin{center}
\caption{Photometric observations\label{tb:lightcurve}}
\tablecaption{Photometric observations\label{tb:lightcurve}}

\begin{tabular}{c c c c c c c c c c c}
\tableline
\tableline
 & & & & \multicolumn{3}{c}{$i_{775}$} & &  \multicolumn{3}{c}{$z_{850}$} \\
\cline{5-7} \cline{9-11}      &       &     &           & Exp. &             &           & & Exp. &             &           \\
Epoch &  Date & MJD & Telescope & (s)  & Scaled Flux & Magnitude & & (s)  & Scaled Flux & Magnitude \\
\tableline
$1$ & 2005-11-28 &  53716.1 & HST &  $175$ &  $0.0018 \pm 0.0049$ & $> 26.515$ &  &  $1400$ &  $-0.0019 \pm 0.0053$ & $> 27.222$ \\
$2$ & 2006-01-03 &  53738.7 & HST &  $375$ &  $0.0002 \pm 0.0025$ & $> 27.509$ &  &  $1500$ &  $0.0053 \pm 0.0049$ &  $26.733 \pm  0.857$ \\
$3$ & 2006-01-29 &  53764.6 & HST & \ldots & \ldots & \ldots &  &  $1500$ &  $0.0087 \pm 0.0050$ &  $26.185 \pm  0.524$ \\
$4$ & 2006-02-21 &  53787.2 & HST &  $515$ &  $0.1183 \pm 0.0032$ &  $23.395 \pm  0.025$ &  &  $1360$ &  $0.1367 \pm 0.0059$ &  $23.201 \pm  0.040$ \\
$5$ & 2006-03-19 &  53813.7 & HST &  $440$ &  $0.4229 \pm 0.0055$ &  $22.012 \pm  0.012$ &  &  $1360$ &  $0.3805 \pm 0.0067$ &  $22.089 \pm  0.016$ \\
$6$ & 2006-04-04 &  53829.6 & HST &  $515$ &  $0.6216 \pm 0.0065$ &  $21.593 \pm  0.010$ &  &  $1360$ &  $0.6055 \pm 0.0074$ &  $21.585 \pm  0.011$ \\
$7$ & 2006-04-22 &  53847.0 & HST &  $515$ &  $0.8343 \pm 0.0068$ &  $21.274 \pm  0.008$ &  &  $1360$ &  $0.8276 \pm 0.0080$ &  $21.246 \pm  0.009$ \\
$8$ & 2006-05-21 &  53876.8 & HST &  $295$ &  $1.0000 \pm 0.0099$ &  $21.077 \pm  0.009$ &  &  $1400$ &  $1.0000 \pm 0.0082$ &  $21.040 \pm  0.008$ \\
$9$ & 2006-06-03 &  53889.3 & HST &  $800$ &  $0.8534 \pm 0.0056$ &  $21.249 \pm  0.006$ &  &  $1200$ &  $0.9176 \pm 0.0081$ &  $21.134 \pm  0.008$ \\
$10$ & 2006-06-28 &  53914.4 & Subaru &  $960$ &  $0.6290 \pm 0.1441$ &  $21.581 \pm  0.211$ &  &  $480$ &  $0.7384 \pm 0.1520$ &  $21.370 \pm  0.189$ \\
$11$ & 2006-08-23 &  53970.3 & Subaru &  $600$ &  $0.0586 \pm 0.0234$ &  $24.158 \pm  0.368$ &  &  $600$ &  $0.0654 \pm 0.0875$ & $> 23.080$ \\
$12$ & 2007-05-18 &  54238.5 & Subaru &  $2280$ &  $-0.0324 \pm 0.0201$ & \ldots &  &  \ldots & \ldots & \ldots \\
\tableline
\end{tabular}

\tablecomments{Flux measurements scaled relative to highest flux epoch; 
               effective zeropoints are 21.077 for $i_{775}$ and 21.040 
	       for $z_{850}$.}
\end{center}
\end{table*}

\begin{figure}
\begin{center}
\epsscale{1.1}
\plotone{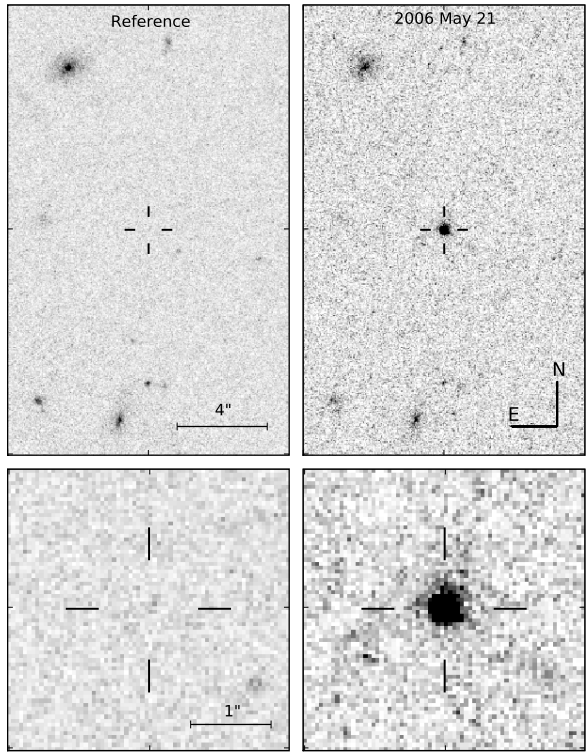}
\end{center}
\caption{Deep stack of the first three epochs in $z_{850}$ totaling 4400~s 
         where the transient is undetected 
	 (\emph{top left} and zoomed in, \emph{bottom left}), 
	 and the highest-flux epoch eight $z_{850}$ exposure of 1400~s 
	 (\emph{top right} and zoomed in, \emph{bottom right}). 
	 All images have the same greyscale.
	 The hash marks indicate the transient position and have the same 
	 physical scale in all images.\label{fig:images}}
\end{figure}

\begin{figure}
\begin{center}
\epsscale{1.2}
\plotone{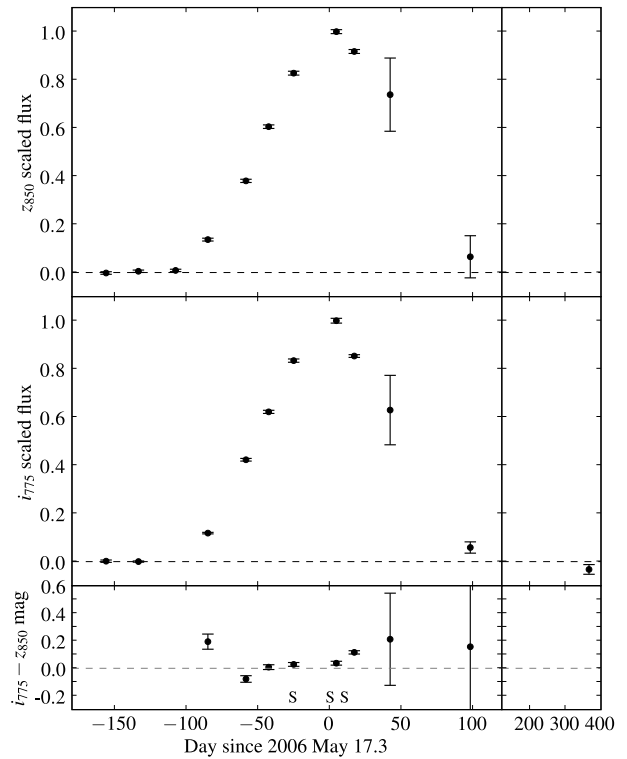}
\end{center}
\caption{Flux lightcurve for $z_{850}$ (\emph{top panel})
         and $i_{775}$ (\emph{middle panel}) scaled to maximum flux. 
	 The last three epochs (starting at +42 days) are 
	 Subaru FOCAS observations. 
	 \emph{bottom panel}: $i_{775} - z_{850}$ color for 
	 epochs with significant detection in both bands. Though the color 
	 only varies $\sim$0.2 magnitudes between the five best measured 
	 epochs, there is evidence for evolution. 
	 The spectral epochs are marked along the abscissa with an ``S.''
	 \label{fig:lightcurve}}
\end{figure}

The transient is consistent with a point source in each of the six ACS 
detection epochs to the extent we can determine.
We performed aperture photometry on drizzled ACS images \citep{fruchter02a} 
using 3.0 pixel ($0.''15$) radius apertures for $i_{775}$ and 5.0~pixel 
($0.''25$) radius apertures for $z_{850}$. Aperture corrections were taken 
from Table~3 of \citet{sirianni05a}. The systematic error due to the known 
color dependence of $z_{850}$ aperture correction \citep[see][]{sirianni05a} 
is estimated to be less than 0.015~mag.

After the transient had left the visibility window of \emph{HST} it remained 
visible from Mauna Kea for several months. Three additional photometry points 
were obtained with the Faint Object Camera and Spectrograph 
\citep[FOCAS;][]{kashikawa02a} 
on the Subaru telescope on 2006 June 28, 2006 August 23, 
and the next year on 2007 May 18. The June observations suffered from poor 
weather conditions ($\mathrm{seeing}\ge 2''$). 
All observations were cosmic ray rejected using 120~s exposures. 
We performed aperture photometry using a $1.''04$ radius aperture and 
estimated photometric errors as described by \citet{morokuma08a}. 
In order to express magnitudes in the ACS filter 
system, we determined Subaru image zeropoints by cross-correlating the 
photometry of nine surrounding stars in the ACS and Subaru images. The 
Subaru FOCAS $i'$ and $z'$ filters are similar enough to ACS $i_{775}$ and 
$z_{850}$ that there is no significant trend with stellar color.

A deep stack of the first three epochs in $z_{850}$ totaling 4400~s 
(Fig.~\ref{fig:images}) and first two epochs 
in $i_{775}$ totaling 550~s provide limits on the magnitude of a possible 
progenitor star (if galactic) or host galaxy (if extragalactic).
No progenitor star is detected in a 3.0~pixel radius aperture centered at the 
position of the transient (known to $< 0.2$~pixels) to a $3\sigma$ upper limit 
of $i_{775} > 26.4$ and $z_{850} > 26.1$ 
(Vega magnitudes are used throughout this paper).
There is no sign of a host galaxy in the 1~arcsec$^2$ surrounding the 
transient to a surface brightness $3\sigma$ limit of 25.0~mag~arcsec$^{-2}$ and 
25.1~mag~arcsec$^{-2}$ in $z_{850}$ and $i_{775}$, respectively.
However, there is a $6\sigma$ detection in a 3.0~pixel radius aperture of a 
$\sim$25.8~mag object $1.''5$ southwest of the 
transient position in $z_{850}$ (Fig.~\ref{fig:images}, \emph{lower left}). 
If the transient is extragalactic, this might represent a faint host galaxy.

The transient increased in brightness in each of 
epochs four through eight before finally declining in the ninth epoch, 
resulting in a rise time of approximately 100~days 
(Fig. \ref{fig:lightcurve}). 
A fit to the brightest five ACS $z_{850}$ photometry points 
gives a date of max of 2006 May 17.3 (MJD 53872.3). 
The declining part of the lightcurve, although sparsely measured, 
is consistent with symmetry about the maximum. 
The final photometry point approximately one year after maximum light shows 
no detection. The $i_{775} - z_{850}$ color is approximately constant over 
the 50~days preceding maximum light, but does show significant signs of 
evolution at early times and after maximum light.

\section{Spectroscopy}

\begin{figure}
\begin{center}
\epsscale{1.1}
\plotone{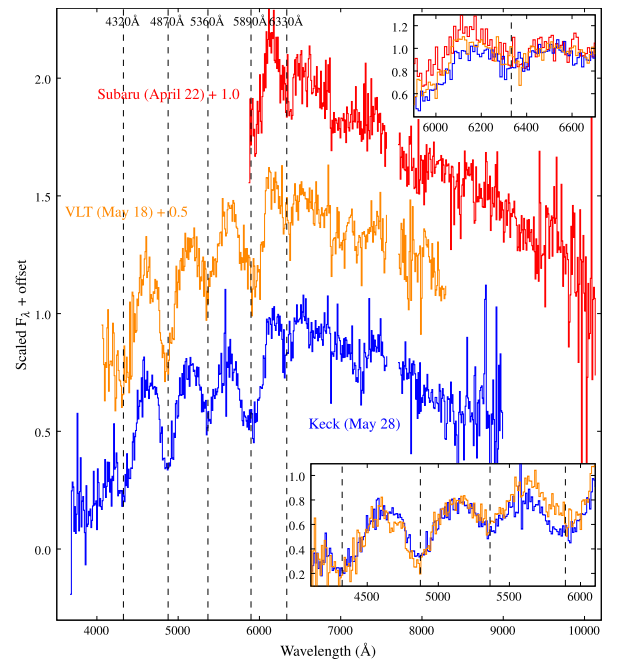}
\end{center}
\caption{Spectra averaged in 10~\AA\ bins. Vertical dotted lines indicate 
         the approximate absorption band centroids. Spectra are normalized 
	 to match in the red continuum. Inset figures show regions where 
	 spectra differ. 
	 \emph{Top Inset}: Overplot of all three spectra (no offset) in the 
	 range 5900 - 6700~\AA, demonstrating apparent evolution of 
	 the flux at $\sim$6150~\AA\ relative to the red continuum.
	 \emph{Bottom Inset}: Overplot of VLT and Keck spectra (no offset) 
	 demonstrating apparent evolution at 4670~\AA\ and of the absorption
	 feature at 5890~\AA.\label{fig:spectra}}
\end{figure}

Spectroscopy was acquired on three dates (Fig.~\ref{fig:spectra}): 
2006 April 22 ($-25$~days) using Subaru FOCAS, 
2006 May 18 ($+1$~day) using VLT FORS2 \citep{appenzeller98a}, and 
2006 May 28 ($+11$~days) with Keck LRIS \citep{oke95a}.\footnote{Spectroscopic 
data is available electronically from 
\texttt{http://supernova.lbl.gov/2006Transient/}}
The Subaru spectrum covers wavelengths longward of 5900~\AA, 
while the VLT and Keck spectra cover bluer wavelengths.
The VLT spectrum (observed at airmass $> 2$) is corrected for differential 
slit loss by applying a linear correction with a slope of 0.25 per 1000~\AA, 
derived from a comparison to the Keck spectrum, which covers the 
entire wavelength range of the VLT spectrum.
The Keck observation was made at the parallactic angle, 
while Subaru FOCAS is equipped with an atmospheric dispersion corrector,
making the Keck and Subaru observations more reliable measures of 
relative flux.

The spectra show a red continuum and several broad absorption features: 
a possible absorption feature at 4320~\AA\ ($\mathrm{FWHM} \sim 180$~\AA), 
three strong features at 4870 ($\mathrm{FWHM} \sim 200$), 
5360 ($\mathrm{FWHM} \sim 230$), and 5890~\AA\ 
($\mathrm{FWHM} \sim 280$~\AA), 
and a weaker absorption feature at 6330~\AA\ ($\mathrm{FWHM} \sim 270$~\AA). 
Errors are estimated to be 15~\AA.

A comparison of the three spectra shows evidence for spectral evolution.
The flux at 
$\sim$6150~\AA\ consistently decreases relative to the red continuum over time
(Fig.~\ref{fig:spectra}, \emph{upper inset}).
Over the 10 day period from the VLT to the Keck spectrum, 
the absorption feature at 5890~\AA\ appears to move toward shorter wavelengths,
while a small absorption feature at 4670~\AA\ in the VLT spectrum 
seems to disappear in the Keck spectrum
(Fig.~\ref{fig:spectra}, \emph{lower inset}).

We compared the spectra to all supernova types using the $\chi^2$ fitting 
program described in \citet{howell05a} as well as the program SNID 
\citep{blondin07a}. No match was found with either program.

Although the cause of the broad absorption features is unclear, we note that 
$\mathrm{d}\lambda/\lambda$ for most of the features is consistent with 
$\sim$0.042 (with the possible exception of the feature at 6330~\AA). 
If the width of the absorption features is due to a velocity distribution, 
the FHWM of this distribution would be $\sim$12,600~km~s$^{-1}$.

One possibility is that the transient is galactic ($z=0$). 
For a galactic source, the slope of the red continuum gives a lower limit 
blackbody temperature of 6500~K.
The absorption features at 4320 and 4870~\AA\ are consistent with
$\mathrm{H}\gamma$ (4341~\AA) and $\mathrm{H}\beta$ (4861~\AA) respectively, 
including uncertainty in the shape of the underlying continuum.
However, there is no significant $\mathrm{H}\alpha$ (6563~\AA) emission or 
absorption, which would be expected for the presence of strong 
$\mathrm{H}\gamma$ and $\mathrm{H}\beta$ features. (Although there is 
slight evidence for emission at 6563~\AA\ in the 
Keck spectrum, this is not seen in the VLT or Subaru spectra.)
The absorption feature at 5890~\AA\ is consistent with Na~\textsc{i} 
$\lambda\lambda$5890, 5896. Although the combination of $\mathrm{H}\gamma$, 
$\mathrm{H}\beta$, and Na~\textsc{i} consistently fits three of the observed 
features, the strong feature at 5360~\AA\ and the weaker feature at
6330\AA\ are left unexplained.
No other narrowband emission or absorption lines are detected.
The spacing of the five absorption features is inconsistent with 
periodicity in energy, as might be expected for cyclotron harmonics. 
Oddly, the features are nearly (but not exactly) periodic in wavelength.

It is also possible that the transient is extragalactic. 
The absence of Lyman $\alpha$ absorption 
features shortward of 4500 \AA\ places a hard upper limit of $z \le 2.7$ on 
its redshift. Among redshifts $0 < z < 2.7$, the cluster redshift of 
$z = 1.112$ is of specific interest as 
the transient is located a small projected distance from the center of the 
cluster. At this redshift,
the absorption feature at 5890~\AA\ is consistent with Mg~\textsc{ii}
$\lambda\lambda$2796, 2803. However, the remaining features are not identified 
at this redshift. 
Given the observed spectral energy distribution, a peak apparent magnitude 
of $i_{775} = 21.077$ implies a total peak observed flux of 
$\sim 2.5 \times 10^{-14}~\mathrm{erg}~\mathrm{s}^{-1}~\mathrm{cm}^{-2}$.
At $z = 1.112$, this implies a peak absolute bolometric magnitude of 
approximately $-22.1$.
This is comparable to the peak absolute magnitudes of SN 2005ap 
\citep[$-22.7$;][]{quimby07a} and SN 2006gy \citep[$-22$;][]{smith07a}, 
the two most luminous SNe observed to date.
Finally, we note that the shape of the continuum is inconsistent with 
$\mathrm{F}_\lambda \propto \lambda^{-5/3}$ synchrotron radiation.

\section{Discussion and Summary}

We have presented photometric and spectroscopic data of an unusual optical 
transient discovered during the \emph{HST} Cluster SN Survey. 
Its key features are as follows:
a rise time of $\sim$100 days with a roughly symmetric lightcurve;
small but statistically significant color variations across the lightcurve;
no detected host galaxy or progenitor; 
broad spectral features in the blue, with a red continuum, and 
some evidence for spectral evolution.
Below, we first discuss constraints on the distance to the source.
Next we consider the possibility that the transient is the result of 
microlensing, finding this to be unlikely. 
Lastly, we search for similar objects in the Sloan Digital Sky Survey (SDSS) 
spectral database, finding no convincing matches.

\begin{figure}
\begin{center}
\epsscale{1.0}
\plotone{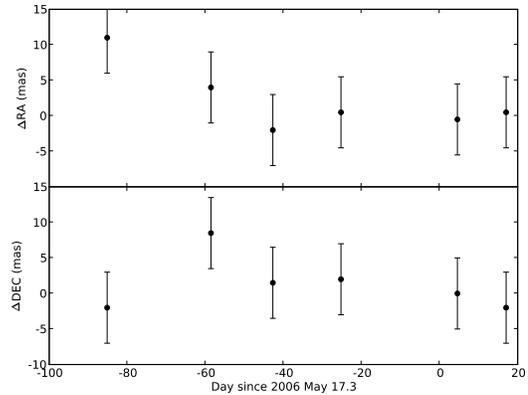}
\end{center}
\caption{Position of transient in each of the six ACS detection epochs 
         with respect to the overall best fit position,
         $\alpha = 14^\mathrm{h} 32^\mathrm{m} 27^\mathrm{s}.395$, 
         $\delta = +33^\circ 32' 24''.83$ (J2000.0). 
         The \emph{top} and \emph{bottom} panels are the position in 
         Right Ascension ($\alpha$) and Declination ($\delta$), respectively.
         5 mas = 0.1 pixels.  
	 \label{fig:propmotion}}
\end{figure}

Any detection of proper motion or parallax would be strong evidence of a 
galactic source. We tested for this by fitting the position of the transient 
in each of the six ACS detection epochs using a 2-d Gaussian 
(Fig.~\ref{fig:propmotion}). 
In all epochs the position uncertainty is dominated by image coalignment 
errors caused by residual distortion, rather than statistical error in the fit.
Note that this uncertainty in relative position between epochs is distinct 
from the uncertainty in the absolute position of the transient discussed 
in \S2. 
For these images and this position, we estimate this error to be 
~0.1 pixel ($0.''005$) in each coordinate. 
The most discrepant positions differ by approximately 0.25~pixels ($0.''0125$).
As a whole, the positions are consistent with no 
proper motion or parallax and give little indication of either. The upper
limit on proper motion is $62~\mathrm{mas}~\mathrm{yr}^{-1}$.
The upper limit on parallax is $\sim$25~mas, which gives
a lower limit on distance of $\sim$40~pc. This limit excludes virtually any 
possible solar system object as the source of the brightening.

We can derive a more significant constraint on distance from reference image
magnitude limits, assuming that the transient is an outburst and a 
progenitor star exists. The dimmest stars 
(aside from neutron stars) known to undergo outbursts are white dwarfs (WDs). 
WDs range in absolute magnitude from approximately 10~mag 
($T_\mathrm{eff}\sim25000$~K) to approximately 16~mag 
($T_\mathrm{eff}\sim3000$~K) and dimmer. 
If we assume the 
progenitor is a WD with absolute magnitude $i_{775} = 16$, the 
reference image $3\sigma$ upper limit of $i_{775} > 26.4$ gives a 
distance $3\sigma$ lower 
limit of $\sim$1.2~kpc. If the 
progenitor is a source dimmer than $i_{775} = 16$
(e.g., a cooler WD or a neutron star), 
the constraints on distance are weaker.
Because the source is at high galactic latitude ($b = 67.3^\circ$), a limit 
of 1.2~kpc places the WD outside the thin disk of the galaxy, as the thin disk 
has a WD scale height of approximately 300~pc \citep{boyle89a}. 
However, there is a significant population of relatively cool WDs 
residing in the thick disk and stellar halo.
Galaxy models predict between tens \citep{castellani02a} and 
hundreds \citep{robin03a} of WDs far dimmer than our reference image 
detection limit in a single ACS field at this galactic latitude. 
Despite the large range (which reflects the uncertainty in the density 
of ancient WDs in the stellar halo) it is clear that the density is high
enough to make a galactic WD a possible progenitor.

Although the symmetry of the lightcurve (Fig.~\ref{fig:lightcurve}) suggests 
that the transient is a microlensing event, this interpretation is unlikely. 
The lightcurve is dramatically broader than the theoretical lightcurve for 
microlensing of a point source by a single lens \citep{paczynski86a}. The 
typical lightcurve FWHM of high-magnification (peak magnification $\ge$ 300) 
microlensing events is on the order of a few hours 
\citep[e.g.,][]{abe04a,dong06a} whereas the transient's lightcurve FWHM 
is $\sim$100 days with a peak magnification $3\sigma$ lower limit of $\sim$120.
Also, the color evolves a small but significant amount over the lightcurve, 
particularly between epochs eight and nine. Some of these difficulties can be 
overcome if we assume the source is resolved; this can both change the 
shape of the lightcurve and allow for color variation as different source 
regions are differentially magnified. However, this typically results in 
a lower peak magnification. Finally, microlensing would still not explain 
the mysterious spectrum.

In an effort to identify objects with similar spectra, we 
cross-correlated the broad features of the spectrum with the SDSS 
spectral database. Each SDSS spectrum was warped with a 
polynomial function to best fit the Keck spectrum, based on a least 
squares fit. The value of the root mean square of the difference between 
the spectra was used to determine the correlation. An allowance for relative
redshift was made, with the requirement that the spectra overlapped in the 
range of the strongest features (3500 to 6200~\AA). No convincing matches 
were found. Changing the warping function between linear and quadratic and 
varying the wavelength range used in the fit altered which
SDSS objects had the highest correlation, but did not result in a more 
convincing match. The SDSS objects with the highest correlation 
were broad absorption line quasars (BAL QSOs) at various redshifts 
and carbon (DQ) WDs.
Although BAL QSOs do have similarly broad features, they don't exhibit 
the spacing or rounded profiles of those of the transient. Also, BAL QSOs 
typically 
include emission features. The DQ WDs most similar to the transient are 
known as DQp WDs. Like the transient, DQp WDs have
broad, rounded absorption features between 4000 and 6000~\AA\ with a red 
continuum \citep[see, e.g.,][]{hall08a}. However, the positions and 
spacing of the 
absorption features shortward of 5000~\AA\ differ greatly from those of the 
transient spectrum. In addition, DQp WDs show increased emission toward the 
UV, which is not seen in the transient. 

The absence of similar spectra in the SDSS database implies that such 
variables are either very rare or typically fainter than the SDSS detection 
threshold, or both. If they are typically faint, this would seem
to argue for an extragalactic origin, though a galactic origin is of course
still possible. 
If this transient does indeed represent a new class of either
galactic or extragalactic transients, such objects will be of great 
interest for future extensive surveys of the time-variable sky.

\acknowledgements
We thank James Graham and Patrick Hall for helpful discussion of spectroscopy.
We thank Ken'ichi Nomoto and Masaomi Tanaka for comparisons to supernovae.
We thank Robert DaSilva for an initial SDSS spectroscopy comparison.
Financial support for this work was provided by NASA through program
GO-10496 from the Space Telescope Science Institute, which is operated by
AURA, Inc., under NASA contract NAS 5-26555. 
This work was also supported in part by the Director, Office of Science, 
Office of High Energy and Nuclear Physics, of the U.S. Department of Energy 
under Contract No. AC02-05CH11231, as well as a JSPS core-to-core program
``International Research Network for Dark Energy'' and by a JSPS research 
grant (20040003).
The authors wish to recognize and acknowledge the very significant cultural
role and reverence that the summit of Mauna Kea has always had within the
indigenous Hawaiian community.  We are most fortunate to have the
opportunity to conduct observations from this mountain.
Finally, this work would not have been possible without the dedicated
efforts of the daytime and nighttime support staff at the Cerro Paranal
Observatory.

{\it Facilities:} \facility{HST (ACS)}, \facility{Subaru (FOCAS)}, 
\facility{Keck:I (LRIS)}, \facility{VLT:Antu (FORS2)}


\begin{thebibliography}{24}

\bibitem[{{Abe} {et~al.}(2004){Abe}, {Bennett}, {Bond}, {Eguchi}, {Furuta},  {Hearnshaw}, {Kamiya}, {Kilmartin}, {Kurata}, {Masuda}, {Matsubara},  {Muraki}, {Noda}, {Okajima}, {Rakich}, {Rattenbury}, {Sako}, {Sekiguchi},  {Sullivan}, {Sumi}, {Tristram}, {Yanagisawa}, {Yock}, {Gal-Yam}, {Lipkin},  {Maoz}, {Ofek}, {Udalski}, {Szewczyk}, {{\.Z}ebru{\'n}}, {Soszy{\'n}ski},  {Szyma{\'n}ski}, {Kubiak}, {Pietrzy{\'n}ski}, \& {Wyrzykowski}}]{abe04a}
{Abe}, F., {et~al.} 2004, Science, 305, 1264

\bibitem[{{Appenzeller} {et~al.}(1998){Appenzeller}, {Fricke}, {F{\"u}rtig},  {G{\"a}ssler}, {H{\"a}fner}, {Harke}, {Hess}, {Hummel}, {J{\"u}rgens},  {Kudritzki}, {Mantel}, {Meisl}, {Muschielok}, {Nicklas}, {Rupprecht},  {Seifert}, {Stahl}, {Szeifert}, \& {Tarantik}}]{appenzeller98a}
{Appenzeller}, I., {et~al.} 1998, The Messenger, 94, 1

\bibitem[{{Blondin} \& {Tonry}(2007)}]{blondin07a}
{Blondin}, S. \& {Tonry}, J.~L. 2007, \apj, 666, 1024

\bibitem[{{Boyle}(1989)}]{boyle89a}
{Boyle}, B.~J. 1989, \mnras, 240, 533

\bibitem[{{Castellani} {et~al.}(2002){Castellani}, {Cignoni}, {Degl'Innocenti}, {Petroni}, \& {Prada Moroni}}]{castellani02a}
{Castellani}, V., {Cignoni}, M., {Degl'Innocenti}, S., {Petroni}, S., \& {Prada Moroni}, P.~G. 2002, \mnras, 334, 69

\bibitem[{{Condon} {et~al.}(1998){Condon}, {Cotton}, {Greisen}, {Yin},  {Perley}, {Taylor}, \& {Broderick}}]{condon98a}
{Condon}, J.~J., {Cotton}, W.~D., {Greisen}, E.~W., {Yin}, Q.~F., {Perley},  R.~A., {Taylor}, G.~B., \& {Broderick}, J.~J. 1998, \aj, 115, 1693

\bibitem[{{Dawson} {et~al.}(2006){Dawson}, {Aldering}, {Barbary}, {Fadeyev},  {Goldhaber}, {Gude}, {Huang}, {Kim}, {Kowalski}, {Kuznetsova}, {Lee},  {Meyers}, {Nugent}, {Perlmutter}, {Rubin}, {Schlegel}, {Spadafora}, {Suzuki},  {Wang}, {Doi}, {Ihara}, {Morokuma}, {Takanashi}, {Tokita}, {Yasuda},  {Lidman}, {Amanullah}, {Goobar}, \& {Stanishev}}]{dawson06a}
{Dawson}, K., {et~al.} 2006, Central Bureau  Electronic Telegrams, 546, 1

\bibitem[{{Dong} {et~al.}(2006){Dong}, {DePoy}, {Gaudi}, {Gould}, {Han},  {Park}, {Pogge}, {Udalski}, {Szewczyk}, {Kubiak}, {Szyma{\'n}ski},  {Pietrzy{\'n}ski}, {Soszy{\'n}ski}, {Wyrzykowski}, \&  {{\.Z}ebru{\'n}}}]{dong06a}
{Dong}, S., {et~al.} 2006, \apj, 642, 842

\bibitem[{{Elston} {et~al.}(2006){Elston}, {Gonzalez}, {McKenzie}, {Brodwin},  {Brown}, {Cardona}, {Dey}, {Dickinson}, {Eisenhardt}, {Jannuzi}, {Lin},  {Mohr}, {Raines}, {Stanford}, \& {Stern}}]{elston06a}
{Elston}, R.~J., {et~al.} 2006, \apj, 639, 816

\bibitem[{{Fruchter} \& {Hook}(2002)}]{fruchter02a}
{Fruchter}, A.~S. \& {Hook}, R.~N. 2002, \pasp, 114, 144

\bibitem[{{Hall} \& {Maxwell}(2008)}]{hall08a}
{Hall}, P.~B. \& {Maxwell}, A.~J. 2008, ArXiv e-prints, 801

\bibitem[{{Howell} {et~al.}(2006){Howell}, {Sullivan}, {Nugent}, {Ellis},  {Conley}, {Le Borgne}, {Carlberg}, {Guy}, {Balam}, {Basa}, {Fouchez}, {Hook},  {Hsiao}, {Neill}, {Pain}, {Perrett}, \& {Pritchet}}]{howell06a}
{Howell}, D.~A., {et~al.} 2006, \nat, 443, 308

\bibitem[{{Howell} {et~al.}(2005){Howell}, {Sullivan}, {Perrett}, {Bronder},  {Hook}, {Astier}, {Aubourg}, {Balam}, {Basa}, {Carlberg}, {Fabbro},  {Fouchez}, {Guy}, {Lafoux}, {Neill}, {Pain}, {Palanque-Delabrouille},  {Pritchet}, {Regnault}, {Rich}, {Taillet}, {Knop}, {McMahon}, {Perlmutter},  \& {Walton}}]{howell05a}
{Howell}, D.~A., {et~al.} 2005, \apj, 634, 1190

\bibitem[{{Kashikawa} {et~al.}(2002){Kashikawa}, {Aoki}, {Asai}, {Ebizuka},  {Inata}, {Iye}, {Kawabata}, {Kosugi}, {Ohyama}, {Okita}, {Ozawa}, {Saito},  {Sasaki}, {Sekiguchi}, {Shimizu}, {Taguchi}, {Takata}, {Yadoumaru}, \&  {Yoshida}}]{kashikawa02a}
{Kashikawa}, N., {et~al.} 2002, \pasj, 54, 819

\bibitem[{{Kenter} {et~al.}(2005){Kenter}, {Murray}, {Forman}, {Jones},  {Green}, {Kochanek}, {Vikhlinin}, {Fabricant}, {Fazio}, {Brand}, {Brown},  {Dey}, {Jannuzi}, {Najita}, {McNamara}, {Shields}, \& {Rieke}}]{kenter05a}
{Kenter}, A., {et~al.} 2005, \apjs, 161, 9

\bibitem[{{Kulkarni} {et~al.}(2007){Kulkarni}, {Ofek}, {Rau}, {Cenko},  {Soderberg}, {Fox}, {Gal-Yam}, {Capak}, {Moon}, {Li}, {Filippenko}, {Egami},  {Kartaltepe}, \& {Sanders}}]{kulkarni07a}
{Kulkarni}, S.~R., {et~al.} 2007,  \nat, 447, 458

\bibitem[{{Morokuma} {et~al.}(2008){Morokuma}, {Doi}, {Yasuda}, {Akiyama},  {Sekiguchi}, {Furusawa}, {Ueda}, {Totani}, {Oda}, {Nagao}, {Kashikawa},  {Murayama}, {Ouchi}, {Watson}, {Richmond}, {Lidman}, {Perlmutter},  {Spadafora}, {Aldering}, {Wang}, {Hook}, \& {Knop}}]{morokuma08a}
{Morokuma}, T., {et~al.} 2008, \apj, 676, 163

\bibitem[{{Ofek} {et~al.}(2008){Ofek}, {Kulkarni}, {Rau}, {Cenko}, {Peng},  {Blakeslee}, {C{\^o}t{\'e}}, {Ferrarese}, {Jord{\'a}n}, {Mei}, {Puzia},  {Bradley}, {Magee}, \& {Bouwens}}]{ofek08a}
{Ofek}, E.~O., {et~al.} 2008,  \apj, 674, 447

\bibitem[{{Oke} {et~al.}(1995){Oke}, {Cohen}, {Carr}, {Cromer}, {Dingizian},  {Harris}, {Labrecque}, {Lucinio}, {Schaal}, {Epps}, \& {Miller}}]{oke95a}
{Oke}, J.~B., {et~al.} 1995, \pasp, 107, 375

\bibitem[{{Paczynski}(1986)}]{paczynski86a}
{Paczynski}, B. 1986, \apj, 304, 1

\bibitem[{{Perlmutter} {et~al.}(1999){Perlmutter}, {Aldering}, {Goldhaber},  {Knop}, {Nugent}, {Castro}, {Deustua}, {Fabbro}, {Goobar}, {Groom}, {Hook},  {Kim}, {Kim}, {Lee}, {Nunes}, {Pain}, {Pennypacker}, {Quimby}, {Lidman},  {Ellis}, {Irwin}, {McMahon}, {Ruiz-Lapuente}, {Walton}, {Schaefer}, {Boyle},  {Filippenko}, {Matheson}, {Fruchter}, {Panagia}, {Newberg}, {Couch}, \& {The  Supernova Cosmology Project}}]{perlmutter99a}
{Perlmutter}, S., {et~al.} 1999, \apj, 517, 565

\bibitem[{{Quimby} {et~al.}(2007){Quimby}, {Aldering}, {Wheeler},  {H{\"o}flich}, {Akerlof}, \& {Rykoff}}]{quimby07a}
{Quimby}, R.~M., {Aldering}, G., {Wheeler}, J.~C., {H{\"o}flich}, P.,  {Akerlof}, C.~W., \& {Rykoff}, E.~S. 2007, \apjl, 668, L99

\bibitem[{{Rau} {et~al.}(2007){Rau}, {Kulkarni}, {Ofek}, \& {Yan}}]{rau07b}
{Rau}, A., {Kulkarni}, S.~R., {Ofek}, E.~O., \& {Yan}, L. 2007, \apj, 659, 1536

\bibitem[{{Robin} {et~al.}(2003){Robin}, {Reyl{\'e}}, {Derri{\`e}re}, \&  {Picaud}}]{robin03a}
{Robin}, A.~C., {Reyl{\'e}}, C., {Derri{\`e}re}, S., \& {Picaud}, S. 2003,  \aap, 409, 523

\bibitem[{{Sirianni} {et~al.}(2005){Sirianni}, {Jee}, {Ben{\'{\i}}tez},  {Blakeslee}, {Martel}, {Meurer}, {Clampin}, {De Marchi}, {Ford}, {Gilliland},  {Hartig}, {Illingworth}, {Mack}, \& {McCann}}]{sirianni05a}
{Sirianni}, M., {et~al.} 2005, \pasp, 117, 1049

\bibitem[{{Smith} {et~al.}(2007){Smith}, {Li}, {Foley}, {Wheeler}, {Pooley},  {Chornock}, {Filippenko}, {Silverman}, {Quimby}, {Bloom}, \&  {Hansen}}]{smith07a}
{Smith}, N., {et~al.} 2007, \apj, 666, 1116

\end{thebibliography}
\end{document}